\documentclass[twocolumn,showpacs,preprintnumbers,amsmath,amssymb,superscriptaddress]{revtex4}

\usepackage{graphicx}
\usepackage{dcolumn}
\usepackage{bm}
\usepackage{color}
\usepackage{psfrag}

\newcommand{\be}{\begin{equation}}
\newcommand{\ee}{\end{equation}}
\newcommand{\beq}{\begin{eqnarray}}
\newcommand{\eeq}{\end{eqnarray}}

\def\nue{\mathrel{{\nu_e}}}
\def\nus{\mathrel{{\nu_s}}}

\def \gta {\mathrel{\vcenter{\hbox{$>$}\nointerlineskip\hbox{$\sim$}}}}

\definecolor{yellow}{rgb}{1,0.9,0} 
\definecolor{wine}{rgb}{0.5,0,0.4}
\definecolor{ancientrose}{rgb}{0.7,0,0.4}
\definecolor{cream}{rgb}{1.,1.,0.7}
\definecolor{violet}{rgb}{1.,0.9,0.95}
\definecolor{lightgreen}{rgb}{0.8,1,0.8}
\definecolor{darkgreen}{rgb}{0.1,0.65,0.1}

\begin{document}

\preprint{LA-UR-03-2162,
         NSF-KITP-03-20}

\title{Neutrino flavor conversion in a neutrino background: single- versus multi-particle  description}

\author{Alexander Friedland}%
 \email{friedland@lanl.gov} \affiliation{Theoretical Division, T-8, MS
 B285, Los Alamos National Laboratory, Los Alamos, NM 87545}
\affiliation{%
School of Natural Sciences, Institute for Advanced Study, Einstein
Drive,
 Princeton, NJ 08540
}%

\author{Cecilia Lunardini}%
 \email{lunardi@ias.edu}
\affiliation{%
School of Natural Sciences, Institute for Advanced Study, Einstein
Drive,
 Princeton, NJ 08540
}%

\date{\today}

\begin{abstract}

In the early Universe, or near a supernova core, neutrino flavor
evolution may be affected by coherent neutrino-neutrino
scattering. We develop a microscopic picture of this phenomenon.
We show that coherent scattering does not lead to the formation of
entangled states in the neutrino ensemble and therefore the
evolution of the system can always be described by a set of
one-particle equations. We also show that the previously accepted
formalism overcounts the neutrino interaction energy; the correct
one-particle evolution equations for both active-active and
active-sterile oscillations contain additional terms. These
additional terms modify the index of refraction of the neutrino
medium, but have no effect on oscillation physics.

\end{abstract}

\maketitle

\section{Introduction}

It is well known that interaction with a medium modifies neutrino
dispersion relations
or, equivalently, gives the medium a nontrivial refraction index
for neutrinos. Because this effect is generically
flavor-dependent, it can have a profound impact on neutrino flavor
evolution. Inside the Sun and the Earth, the refraction effect
arises as a result of neutrino interactions with  electrons and
nucleons. On the other hand, in the early Universe or near a
supernova core, where the number density of neutrinos is
sufficiently high, the refractive properties of the neutrino
background itself are important
\cite{SmirnovMoriond,SchrammFuller1987,NotzoldRaffelt1988,BarbieriDolgov1990,SavageFuller}.
This neutrino ``self-refraction" is the subject of the present
work.

Early studies
\cite{SchrammFuller1987,NotzoldRaffelt1988,BarbieriDolgov1990,SavageFuller,BarbieriDolgov1991,Langacker:1992xk}
of the effect treated the neutrino background analogously to the
case of ordinary matter (electrons and nucleons), namely, assuming
that
\begin{itemize}
    \item for each neutrino, one can write a single-particle
evolution equation and take into account the effect of all other
particles (including other neutrinos) by adding appropriate terms
to the one-particle Hamiltonian;
    \item the contribution of the background
    neutrinos to this Hamiltonian is diagonal in the flavor basis.
\end{itemize}
With these assumptions, the evolution of each neutrino in the
system was described by
\begin{eqnarray}
\label{eq:Pant_tot}
 i\frac{d\nu^{(i)}}{dt}=(H_{\rm vac}+H_{\rm mat}+\sum_j H_{\nu\nu}^{(ij)}) \nu^{(i)},
\end{eqnarray}
where $i$ and $j$ label neutrinos in the system, $H_{\rm vac}$ and
$H_{\rm mat}$ are the usual vacuum and ordinary matter Hamiltonian
terms and the last term is the sum over the contributions of all
background neutrinos, taken to be flavor-diagonal.

These assumptions were critically reexamined by James Pantaleone
\cite{Pantaleone1992prd,Pantaleone1992plb} who made several
important observations regarding their validity. First, he
reasoned that flavor evolution of neutrinos in a neutrino
background is in general a many-body phenomenon and it is {\it a
priori} not obvious that the first assumption is justified in all
cases. Second, he demonstrated that, for a system of several
active neutrinos flavors, even when the first assumption is
justified, the second one is incompatible with the symmetry of the
problem. Indeed, the low energy neutral current Hamiltonian
\footnote{For physical applications, both in the case of the early
Universe and supernova, the relevant neutrino energies are in the
$10^6-10^7$ eV range.},
\begin{eqnarray}
  \label{eq:NChamiltonian}
  H_{\rm NC}= \frac{G_F}{\sqrt{2}}
\left(\sum_a j_a^\mu\right)\left(\sum_b j_{b\mu}\right),
\end{eqnarray}
where the currents $j_a^\mu\equiv \bar\nu_a\gamma^\mu\nu_a$ and
$a$ is a flavor index, $a=1,..N$, possesses a $U(N)$ flavor
symmetry, which must be respected by any effective description of
the system. This requirement is not satisfied by the diagonal form
for $H_{\nu\nu}^{(ij)}$ used in the earlier studies.

Pantaleone proposed a modified form for
$H_{\nu\nu}^{(ij)}$, which contained non-zero off-diagonal terms,
\begin{equation}
\label{eq:Pantaleone}
H_{\nu\nu}^{(ij)}=A\left[|\nu_e^{(j)}|^2+|\nu_\mu^{(j)}|^2+
\begin{pmatrix}
  |\nu_e^{(j)}|^2 & \nu_e^{(j)} \nu_\mu^{(j)\ast} \\
  \nu_e^{(j)\ast} \nu_\mu^{(j)} & |\nu_\mu^{(j)}|^2 \\
\end{pmatrix}\right].
\end{equation}
In Eq.~(\ref{eq:Pantaleone}) the wave function of the background
neutrino $\nu^{(j)}$ is normalized such that $\int dV
(|\nu_e^{(j)}|^2+|\nu_\mu^{(j)}|^2)=1$ and the coefficient of
proportionality $A$ equals $\sqrt{2} G_F (1-\cos\beta^{(ij)})$,
$\beta^{(ij)}$ being the angle between the two neutrino momenta.
For antineutrinos in the background, the form of the Hamiltonian
is exactly the same, with the only difference that $A$ has the
opposite sign. As can be easily checked, Eq.~(\ref{eq:Pantaleone})
is indeed consistent with the $U(2)$ flavor symmetry of the
two-flavor system.

The result (\ref{eq:Pantaleone}) was also later obtained by Sigl
and Raffelt \cite{germanpaper} and by McKellar and Thomson
\cite{McKellarThomson} in the framework of more general analyses
of the flavor evolution of a neutrino system, which took into
account both refraction effects and collisions.
Eqs.~(\ref{eq:Pant_tot},\ref{eq:Pantaleone}) have been used as a
starting point for extensive studies of the neutrino evolution in
the early Universe
\cite{Kostelecky:1993yt,Kostelecky:1993dm,Kostelecky:1994ys,Kostelecky:1995xc,Lunardini2000,Hannestad:2001iy,Dolgovetal2002, Wong2002, Abazajian2002} and in supernovae
\cite{Pantaleone1995,QianFuller1995,Sigl1995,McLaughlin1999,PastorRaffelt2002}.
The general properties of the solutions have been investigated in
\cite{Samuel:1993uw,Samuel:1996ri,Pantaleone:1998xi,Pastor:2001iu}.

Certain theoretical aspects of these equations, however, have not
received adequate attention in the literature to date. One such
aspect is the validity of a description of the problem by a set of
single-particle equations. The absence of quantum correlations
(entanglement) at all times during the evolution of the system is
{\it a priori} not obvious, and has not been proven. Indeed, in
the derivations \cite{germanpaper} and \cite{McKellarThomson} it
is stated as an assumption. In \cite{Pantaleone1992plb}, it is
suggested that for quantum correlations not to form the system
must obey certain physical conditions, for example, it should
contain an incoherent mixture of mass eigenstates. This raises the
question how general the results in
Eqs.~(\ref{eq:Pant_tot},\ref{eq:Pantaleone}) are and what physical
criteria determine their breakdown.

The second important aspect is the connection between the
elementary neutrino-neutrino scattering processes and the
macroscopic description of refraction given in
Eqs.~(\ref{eq:Pant_tot},\ref{eq:Pantaleone}). For refraction in
ordinary matter, such a connection is very well established and
provides the most straightforward derivation of the effect as an
interference of many elementary scattering amplitudes.  A similar
treatment for the case of the neutrino background has not been
given.

In this paper we present a description of the flavor evolution of
neutrinos in a neutrino background in terms of the elementary
neutrino-neutrino scattering processes. This description provides
a transparent, physical picture of the effect, and, at the same
time, allows us to prove that quantum correlations in the neutrino
ensemble are negligible and therefore the description in terms of
one-particle equations is valid. These equations are obtained
directly from our formalism and are compared to the accepted
results, Eqs.~(\ref{eq:Pant_tot},\ref{eq:Pantaleone}).

We work in the regime in which (i) neutrino-neutrino interactions
are described by the low-energy four-fermion Fermi coupling, (ii)
the neutrino gas is non-degenerate, and (iii) incoherent
scattering of neutrinos with other neutrinos and other particle
species is negligible. The first condition implies that the
neutrino center-of-mass energies are well below the weak scale.
The second one is satisfied if a gas of neutrinos has a number
density $n_\nu \ll E^3$, where $E$ is a typical neutrino energy.
Finally, incoherent scattering is significant if the column
density of the medium exceeds the inverse of the scattering cross
section: $d\equiv \int n(x) dx \gta 1/\sigma$, as happens, e.g.,
in the early Universe at temperatures larger than few MeV (see,
e.g., \cite{Dolgovetal2002}).

The refraction effects of the neutrino background are negligible
if the coupling between a neutrino and the neutrino gas is
significantly smaller than the vacuum oscillation Hamiltonian or
than the coupling to ordinary matter. This condition thus depends
on the neutrino energy, on the density of the neutrino background,
and on the density of the ordinary matter.  In the core of the
Sun, where the number density of neutrinos is $n_\nu \simeq
10^{6}~{\rm cm^{-3}}$, neutrino-neutrino interaction is negligible
for all relevant neutrino energies and oscillation parameters. In
contrast, near a supernova core the neutrino density can be as
high as $n_\nu \simeq 10^{31}~{\rm cm^{-3}}$ and is comparable
with the electron density there. Therefore, neutrino-induced
refraction effects are important and must be taken into account.

The paper is organized as follows. In Sect.~\ref{sect:questions} we
review the microscopical picture of the neutrino flavor evolution in
normal matter and point out that a naive extension of this picture to
the neutrino self-refraction leads to seemingly paradoxical
results. In Sect.~\ref{sect:exchanges} we show how this picture should
be constructed consistently by identifying states in the neutrino
ensemble that amplify coherently. We also show that coherent
scattering does not form entangled states. In
Sect.~\ref{sect:comparison} we compare the one-particle evolution
equations we obtain for active-active and active-sterile oscillations
in the neutrino and antineutrino backgrounds with the accepted
results.  In Sect.~\ref{sect:moreent} we show how entanglement is
effectively destroyed by the refraction phenomenon. Finally,
Sect.~\ref{sect:conclusions} summarizes our conclusions.

\section{Formulation of the problem}
\label{sect:questions}

\subsection{Conventional neutrino refraction: FCNC case}
\label{sec:fcnc}

\begin{figure}
\psfrag{nu_e}{\color{blue}$\nu_e$}
\psfrag{nu_mu}{\color{blue}$\nu_\mu$}
  \includegraphics[width=0.45\textwidth]{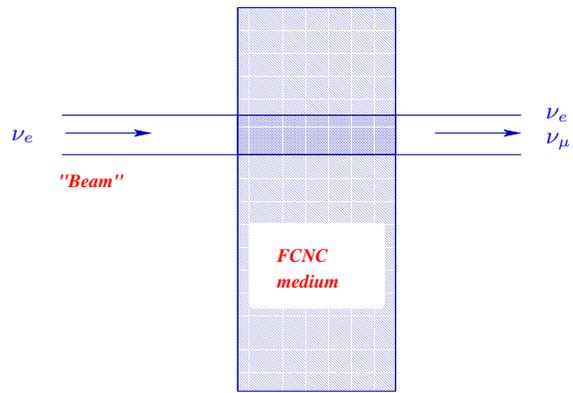}\\
  \caption{Neutrino beam traversing a slab of FCNC interacting matter.}\label{fig:fig2}
\end{figure}

We begin by briefly reviewing the physics of the refraction effect
in ``normal" matter. While there are many ways to derive the
refraction properties in this case, fundamentally, the relevant
effect is the coherent interference of many elementary scattering
events.

Let us consider the problem of neutrino oscillations in a medium
which possesses flavor changing neutral current (FCNC)
interactions. Such interactions give rise to the nonzero
off-diagonal terms in the neutrino evolution Hamiltonian,
\begin{equation}
\label{eq:H_FCNC } H_{FCNC} = \frac{\sqrt{2} G_F
n_2}{2}\left[const+
\begin{pmatrix}
  \epsilon' & \epsilon \\
  \epsilon & -\epsilon' \\
\end{pmatrix}\right],
\end{equation}
where $G_F$ is the Fermi constant and $n_2$ is the number density
of scatterers in the medium.

As a toy example, consider a beam of electron neutrinos incident
on a thin slab of matter of thickness $L$ made of FCNC interacting
particles, as illustrated in Fig.~\ref{fig:fig2}.
Assume that the neutrino masses are sufficiently small so that the
effects of vacuum oscillation can be neglected. The flavor
conversion rate in the slab can then be found using the following
straightforward physical argument. Let $f$ be the amplitude for an
electron neutrino to scatter as a muon neutrino in a given
direction on a particle in the target. If the scattering
amplitudes for different target particles add up incoherently, the
flux of muon neutrinos in that direction is $\propto N_s |f|^2$,
where $N_{s}$ is the number of scatterers. In the case of forward
scattering, however, the scattering amplitudes add up
\emph{coherently} and, hence, the forward flux of muon neutrinos
is $\propto N_s^2 |f|^2$.  Indeed, in the small $L$ limit
Eq.~(\ref{eq:H_FCNC }) gives
\begin{equation}
\label{eq:PnunuFCNCapprox}
P_{\nu_e\rightarrow\nu_\mu}^{FCNC}\simeq \epsilon^2 (G_F n_2
L)^2/2~,
\end{equation}
which has the form $P_{\nu_e\rightarrow\nu_\mu}^{FCNC}\propto
N_s^2 |f|^2$, since $\epsilon\propto f$. Notice that by choosing a
small $L$ limit we were able to ignore the secondary conversion
effects in the slab, {\it i.e.}, to assume that for all elementary
scattering events the incident neutrinos are in the $\nu_e$ state.

To summarize, for small enough $L$, the flavor conversion rate due
to coherent FC scattering in the forward direction is proportional
to the square of the modulus of the product of the elementary
scattering amplitude and number of scatterers. This quadratic
dependence on $N_s$ is what makes the coherent forward scattering
important even when the incoherent scattering can be neglected.

Notice that exactly the same arguments apply if one considers the
usual flavor-diagonal matter term due to the electron background
in a rotated basis, for instance, in the basis of vacuum mass
eigenstates. In this basis, the matter Hamiltonian has
off-diagonal terms, resulting in transitions between the vacuum
mass eigenstates.

\subsection{Neutrino background: physical introduction}

\label{sec:2b}

\begin{figure}
\psfrag{nu_e}{\color{blue}$\nu_e$}
\psfrag{nu_mu}{\color{blue}$\nu_\mu$}
\psfrag{nu_x}{\color{blue}$\nu_x=\cos\alpha \nu_e+\sin\alpha
\nu_\mu$}
  \includegraphics[width=0.5\textwidth]{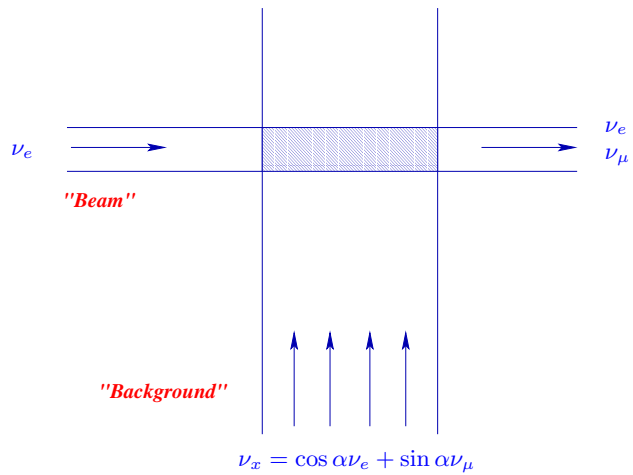}\\
  \caption{Toy problem to illustrate neutrino flavor conversion in the neutrino background. }
  \label{fig:fig1}
\end{figure}

We seek the same description for the case of neutrino background.
Let us therefore modify the setup in Fig.~\ref{fig:fig2} and
replace the slab by a second neutrino beam, such that the neutrino
momenta in the two beams are orthogonal (see Fig.~\ref{fig:fig1}).
To keep the parallel between this case and the FCNC case, we will
continue to refer to the original beam as ``the beam'' and to the
second beam as ``the background''. The neutrinos in each beam can
be taken to be approximately monoenergetic \footnote{To avoid
conflict with the Pauli's principle
and to make possible labelling of neutrinos by their momenta, we can
assume that neutrinos have slightly different energies.}.
We again assume that the neutrino masses are sufficiently
small so that, although flavor superposition states could be created
\emph{outside} the intersection region, the effects of vacuum
oscillation \emph{inside} the intersection region can be
neglected. Any flavor conversion that takes place in the system is
therefore due to neutrino-neutrino interactions in the intersection
region.

\begin{figure}[t]
\psfrag{nu_e}{\large\color{blue}$\nu_e$}
\psfrag{nu_x}{\large\color{blue}$\nu_x$}
  \includegraphics[width=0.5\textwidth]{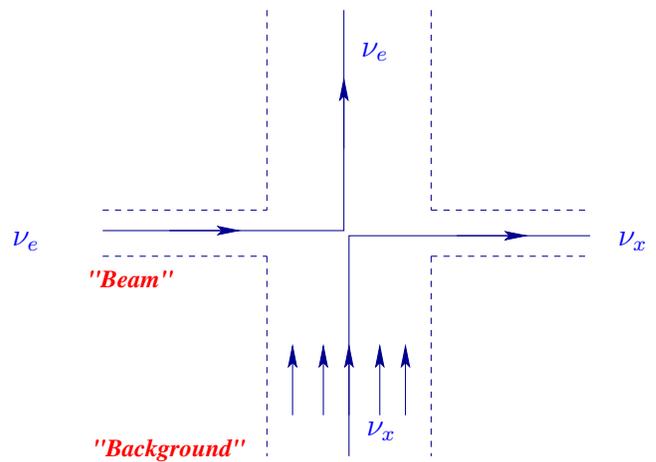}
  \caption{Elementary scattering event that causes a change of the flavor composition of the beam}
  \label{fig:elementary}
\end{figure}

Let us first compute the amount of flavor conversion in the beam
using Eqs.~(\ref{eq:Pant_tot},\ref{eq:Pantaleone}). The conversion
is expected because of the presence of the off-diagonal terms in
these equations. The result depends on the flavor composition of
the background. If the background neutrinos are all in the same
flavor state
\begin{equation}\label{eq:definenux}
    \nu_x=\cos\alpha\nu_e+\sin\alpha\nu_\mu
\end{equation}
and their density is $n_2$, the Hamiltonian for the evolution of a
beam neutrino takes the form
\begin{eqnarray}
\label{eq:Pant_brief}
  H = \frac{\sqrt{2} G_F n_2}{2}\left[const+
\begin{pmatrix}
  \cos 2\alpha & \sin 2\alpha \\
  \sin 2\alpha & -\cos 2\alpha \\
\end{pmatrix}\right].
\end{eqnarray}

After traversing the intersection region, a neutrino in the beam
will be converted to the $\nu_\mu$ state with the probability
\begin{equation}
\label{eq:Pnunuapprox} P_{\nu_e\rightarrow\nu_\mu}\simeq \sin^2
2\alpha (G_F n_2 L)^2/2,
\end{equation}
assuming as before that the size of the region $L$ is small.

The above descriptions of neutrino flavor conversion in the
neutrino background and in the FCNC background are very similar in
form. We will next show, however, that despite very similar
appearances of the equations, the underlying physics in the two
cases is different.

First, we need to establish which elementary processes give rise to
neutrino flavor conversion in the neutrino background. In the
absence of vacuum oscillations, the only interactions in the
problem are neutral current interactions between pairs of
neutrinos. These are described by the Hamiltonian
(\ref{eq:NChamiltonian}), which conserves the \emph{total} flavor
of the system.
The flavor composition of the beam therefore changes only if some
of the background neutrinos of different flavor scatter into the
beam, {\it i.e.}, if a neutrino from the background and a neutrino
from the beam \emph{exchange} momenta,
$\nu(\vec{k})+\nu(\vec{p})\rightarrow\nu(\vec{p})+\nu(\vec{k})$.
As observed in \cite{Pantaleone1992plb}, such events can add up
coherently and give rise to the flavor off-diagonal entries in the
oscillation Hamiltonian.

Let us consider such an elementary event, as depicted in
Fig.~\ref{fig:elementary}. Following \cite{Pantaleone1992plb}, we
restrict scattering angles to the directions in which coherent
interference occurs and write the Hamiltonian for an interacting
pair of neutrinos in the form
\begin{widetext}
\begin{eqnarray}
  \label{eq:pant4by4}
  i\frac{d}{dt}\left(
    \begin{array}{c}
      |\nu_e (\vec{k}) \nu_e(\vec{p})\rangle  \\
      |\nu_e (\vec{k}) \nu_\mu(\vec{p})\rangle \\
      |\nu_\mu (\vec{k}) \nu_e(\vec{p})\rangle \\
      |\nu_\mu (\vec{k}) \nu_\mu(\vec{p})\rangle
    \end{array}\right)
=\frac{\sqrt{2}G_F}{V} (1-\cos\beta) \left(
\begin{array}{cccc}
  2 & 0 & 0 & 0 \\
  0 & 1 & 1 & 0 \\
  0 & 1 & 1 & 0 \\
  0 & 0 & 0 & 2
\end{array}
\right) \left(
    \begin{array}{c}
      |\nu_e (\vec{k}) \nu_e(\vec{p})\rangle  \\
      |\nu_e (\vec{k}) \nu_\mu(\vec{p})\rangle \\
      |\nu_\mu (\vec{k}) \nu_e(\vec{p})\rangle \\
      |\nu_\mu (\vec{k}) \nu_\mu(\vec{p})\rangle
    \end{array}\right),
\end{eqnarray}
\end{widetext}
where $V$ is the normalization volume. For a state $|S\rangle$,
which at $t=0$ is $|\nu_e\nu_\mu\rangle$, this equation formally
has a solution
\begin{equation}
\label{eq:Pant_entangle}
|S(t)\rangle=\frac{|\nu_e\nu_\mu\rangle-|\nu_\mu\nu_e\rangle}{2}+e^{-i\delta E t}\frac{|\nu_e\nu_\mu\rangle+|\nu_\mu\nu_e\rangle}{2},
\end{equation}
where $\delta E=2\sqrt{2}G_F(1-\cos\beta)/V$. In practice,
however, one only needs a small $t$ expansion of Eq. (\ref{eq:Pant_entangle}):
\begin{eqnarray}
\label{eq:small_t}
S(t)&\simeq &
|\nu_e\nu_\mu\rangle-i\delta E t/2(|\nu_e\nu_\mu\rangle+|\nu_\mu\nu_e\rangle) \nonumber\\
&=& (1+i a)|\nu_e\nu_\mu\rangle+i a|\nu_\mu\nu_e\rangle,
\end{eqnarray}
where $a=-\sqrt{2}G_F(1-\cos\beta)t/V$. Indeed, since the
interaction time and the normalization volume are determined by
the size of the neutrino wave packet $l$, $t\sim l$ and $V\sim
l^3$, the absolute value of $a$ in any realistic physical
situation is always much less than 1. This simply means that the
interaction between two neutrinos is described by the lowest order
Feynman diagram and second order scattering effects may be safely
ignored.

We now return to the situation depicted in Fig.~\ref{fig:fig1}.
We are interested in the muon neutrino content of the beam after
it crosses the interaction region. Using the Hamiltonian in
Eq.~(\ref{eq:pant4by4}) to describe an elementary scattering event
$\nu_e+\nu_x$, we find
\begin{eqnarray}
\label{eq:elementary}
 \tilde S(t)&\simeq &
 |\nu_e\nu_x\rangle+i a(|\nu_x\nu_e\rangle+|\nu_e\nu_x\rangle),\nonumber\\
&=& (1+i a)|\nu_e\nu_x\rangle+i a\cos\alpha|\nu_e\nu_e\rangle\nonumber\\
&+&i a\sin\alpha|\nu_\mu\nu_e\rangle.
\end{eqnarray}
Thus, for an elementary event, the amplitude of measuring the
neutrino in the beam as $\nu_\mu$ is proportional to $\sin\alpha$.
If one multiplies this amplitude by the number of scatterers and
squares, as was previously done in the FCNC example, one finds
$P_{\nu_e\rightarrow\nu_\mu} \propto N_s^2 \sin^2\alpha$, which is
in clear conflict with the prediction of
Eq.~(\ref{eq:Pnunuapprox}), $P_{\nu_e\rightarrow\nu_\mu} \propto
N_s^2\sin^2 2\alpha$.

Which of the two results is right? At first sight, the first
possibility may appear more plausible. Consider, for instance,
what happens when an electron neutrino propagates through a
background of muon neutrinos ($\alpha=\pi/2$). Since muon
neutrinos appear in the beam as a result of elementary exchanges
between the beam and the background, one may think that for a pure
muon neutrino background the conversion rate in the beam should be
maximal. The first result indeed has this behavior, while the
second result predicts \emph{no conversion}. One may therefore be
tempted to conclude that the second result fails to describe the
system in this limit and perhaps that this failure signals the
general breakdown of the validity of the single-particle
description.

As will be shown in the next section, however, despite its
seemingly paradoxical behavior, the second result is actually
correct. The explanation of the paradox lies in the procedure of
adding up elementary amplitudes, which in the case of the neutrino
background is more subtle than in the case of ordinary matter.

\section{Microscopic analysis of the effect}
\label{sect:exchanges}

The key observation is that a neutrino-neutrino scattering event
changes not only the state of the beam, but also the state of the
background.  This implies that the refraction by the neutrino
background is intrinsically different from the case of ordinary
matter, and requires a specific analysis to establish what states
in the neutrino ensemble are amplified coherently.

We begin by recalling that an elementary scattering event can only
be amplified if the particles scatter ``forward", meaning that
either particle momenta do not change ($t$-channel), or that they
are exchanged ($u$-channel) as in Fig.~\ref{fig:elementary}. The
interaction is described by Eq.~(\ref{eq:small_t}). For the sake
of clarity, in this section we consider the exchange diagrams only
and omit the effects of the non-exchange diagrams. The latter produce
an overall (flavor-independent) phase shift of the neutrino
states, and do not affect the neutrino flavor conversion.

We first consider a single electron neutrino in the beam
interacting with several neutrinos in the background. Assume that
these background neutrinos are all in the same flavor state
$\nu_x$, Eq. (\ref{eq:definenux}), and that the state of the
background initially is a product of single-particle states, $|x x
x...x x\rangle$. The total system therefore is initially in the
state $|e\rangle |x x x...x x\rangle$ and, as a result of the
interaction, evolves over a time step $\delta t$ according to
\begin{eqnarray}
\label{eq:1nuexch}
  |e\rangle |x x x...x x\rangle
\longrightarrow
| F \rangle=  |e\rangle |x x x...x x\rangle
+ i a |F_1\rangle,
\end{eqnarray}
where $|F_1\rangle$ is the state with all possible exchange terms,
\begin{eqnarray}
 |F_1\rangle &=& |x\rangle
  |e x x...x x \rangle
+ |x \rangle |x e x ...x x \rangle \nonumber\\
&+&|x \rangle |x x e ...x x \rangle + ...\; .
\end{eqnarray}

To measure the $\nu_\mu$ content of the beam, we introduce a
``$\nu_\mu$ number" operator $\hat \mu \equiv
|\mu\rangle\langle\mu|$. This operator acts only on the beam
states and gives $\langle x | \hat \mu|x \rangle=\sin^2\alpha$. To
find the probability that the beam neutrino after interacting with
$N_2$ background neutrinos will be measured as $\nu_\mu$, we
compute the expectation value of $\hat \mu$ in the final state of
Eq.~(\ref{eq:1nuexch})
\begin{eqnarray}
\label{eq:N2matr}
 \langle F| \hat \mu|F\rangle&=&  a^2\langle x | \hat \mu|x \rangle \nonumber\\
&\times&
 (\langle e x x ...x |+
\langle x e x ...x |+
... )\nonumber\\
&& (|e x x ...x \rangle+ |x e x ...x \rangle +... )\nonumber\\
&=&a^2 \sin^2 \alpha ((N_2^2-N_2) \cos^2\alpha + N_2).
\end{eqnarray}
The last line is obtained by observing that in the sum there are
$N_2$ ``diagonal"  terms of the type $\langle x... x e x ...x |
x...x e x ...x\rangle=1$ and $N_2^2-N_2$ ``off-diagonal" terms of
the type $\langle x... x e x ...x |  x... e x x
...x\rangle=\cos^2\alpha$.

The result in Eq.~(\ref{eq:N2matr}) shows how the connection
between a single scattering event and Eq.~(\ref{eq:Pnunuapprox})
is made. For a single scattering event, $N_2=1$, one indeed finds
that the conversion probability
$P_{\nu_e\rightarrow\nu_\mu}\propto\sin^2\alpha$ is maximized when
the background consists of muon neutrinos. As the number of
scattering events increases, however, the maximum of conversion
efficiency shifts to values of $\alpha <\pi/2$. In the limit of
large $N_2$, one finds $P_{\nu_e\rightarrow\nu_\mu}\propto
N_2^2\sin^2 2\alpha$, precisely as predicted by
Eq.~(\ref{eq:Pnunuapprox}).

We can now understand what happens to an electron neutrino
propagating in a muon neutrino background. Eq.~(\ref{eq:N2matr})
shows that the conversion rate does not strictly vanish, as
predicted by Eq.~(\ref{eq:Pnunuapprox}), but  is only proportional
to $N_2$, not to $N_2^2$. As discussed before, the proportionality
to $N_2$ is a feature of incoherent scattering. The coherent
scattering part is indeed strictly zero in this case: the states
$| e \mu \mu \mu ...\rangle$ , $| \mu e \mu \mu ...\rangle$, etc.,
are \emph{mutually orthogonal}, and hence have no overlap that
could be coherently amplified.

It can be readily seen that the part of the final state $| F
\rangle$ that gets amplified contains the projection of the final
background states onto the \emph{initial} background state $|x x x
... \rangle$. Decomposing the states in $|F_1\rangle$ according to
\begin{eqnarray}
\label{eq:project_bg}
|  x...x e x ...\rangle=\langle x| e\rangle|  x...x x x ...\rangle+
 \langle y| e\rangle|  x...x y x ...\rangle,
\end{eqnarray}
where $y$ is the state orthogonal to $x$, $\langle y| x\rangle =
0$, we can write the result (\ref{eq:N2matr}) as
\begin{eqnarray}
\label{eq:N2matr2}
 \langle F| \hat \mu|F\rangle&=&  a^2\langle x | \hat \mu|x \rangle \nonumber\\
&\times&
 (\langle e|x\rangle N_2\langle x x x ...x| +\nonumber\\
 &&\langle e|y\rangle\langle y x x ...x |+
\langle e|y\rangle\langle x y x ...x |+
... )\nonumber\\
&& (\langle x|e\rangle N_2 | x x x ...x\rangle+\nonumber\\
 &&\langle y|e\rangle|y x x ...x \rangle+ \langle y|e\rangle |x y x ...x \rangle +... )\nonumber\\
&=& a^2\langle x | \hat \mu|x \rangle (N_2^2 |\langle x|e\rangle
|^2 + N_2 |\langle y|e\rangle |^2).
\end{eqnarray}

We now return to the problem of two intersecting beams, one
containing electron neutrinos and another containing neutrinos in
the flavor superposition state $\nu_x$. Suppose $N_1$ neutrinos
from the first beam interact with $N_2$ neutrinos from the second
beam. As a result of the interaction, the system evolves over time
$\delta t$ according to
\begin{eqnarray}
\label{eq:2b_exch}
  |e e e ...\rangle
  |x x x ...\rangle
\longrightarrow |F\rangle=
  |e e e ...\rangle
  |x x x ...\rangle
+ i a |F_1\rangle,
\end{eqnarray}
where $|F_1\rangle$ contains a sum of terms with all possible
exchanges,
\begin{eqnarray}
 |F_1\rangle &=&
(|x e e ...e\rangle+|e x e ...e\rangle+|e e x  ...e\rangle+...)\nonumber\\
&\times&(|e x x ...x\rangle+|x e  x ...x\rangle+|x x  e ...x\rangle+...).\;\;\;\;
\end{eqnarray}

As in the derivation of Eq.~(\ref{eq:N2matr2}), to separate the
coherent and incoherent parts, the states in $ |F_1\rangle$ need
to be projected on the corresponding initial states and orthogonal
states. This is done using Eq.~(\ref{eq:project_bg}) and
\begin{eqnarray}
\label{eq:project_beam}
|x e e ...\rangle =\langle e |x \rangle|e e e ...\rangle
+\langle\mu|x \rangle|\mu e e ...\rangle.
\end{eqnarray}
As a result, we obtain
\begin{widetext}
\begin{eqnarray}
  \label{eq:2beams2}
 |F_1\rangle &=&
N_1 N_2 |\langle e |x \rangle|^2  |e e e ...\rangle
|x x x ...\rangle \nonumber\\
&+& N_2 \langle\mu |x \rangle \langle x |e \rangle (|\mu e e
...\rangle+ |e \mu e ...\rangle+...)
|x x x ...\rangle\nonumber\\
&+& N_1 \langle e |x \rangle \langle y  |e \rangle |e e e
...\rangle (|y  x x ...\rangle+ |x y  x ...\rangle+...)
\nonumber\\
&+& \langle\mu |x \rangle \langle y  |e \rangle
 (|\mu e e ...\rangle+|e \mu e ...\rangle+...)
 (|y  x x ...\rangle+|x y  x ...\rangle+...).
\end{eqnarray}
\end{widetext}

We observe that the term in the last line contains a sum of
mutually orthogonal states. This term represents the part that
will not amplify coherently, as can be seen by repeating the
arguments given in connection with Eq.~(\ref{eq:N2matr2}). We
further note that if one \emph{drops this term}, at first order in
$a$ the final state equals
\begin{eqnarray}
  \label{eq:2beams3}
|F\rangle=
  |e' e' e' ... e' e'\rangle
 |x' x' x' ... x' x'\rangle,
\end{eqnarray}
where
\begin{widetext}
\begin{eqnarray}
\label{eq:single_rotated_beam}
  |e'\rangle &=&
  |e \rangle + i N_2 a [1/2\times |\langle x | e \rangle|^2
|e \rangle+ \langle \mu| x \rangle \langle x  |e \rangle
|\mu\rangle],\;\;\;\\
\label{eq:single_rotated_background}
 |x'\rangle &=&
  |x \rangle+i N_1 a[1/2 \times  |\langle e |x \rangle|^2 |x \rangle +
  \langle e |x \rangle \langle y  |e \rangle |y \rangle].\;\;\;
\end{eqnarray}
\end{widetext}

Eqs.~(\ref{eq:2beams3},\ref{eq:single_rotated_beam},\ref{eq:single_rotated_background})
represent the central result of this section. They show that if we
take the initial state to be a product of single-particle neutrino
states and evolve it in time --- carefully separating the coherent
effects and dropping the incoherent ones --- the final state
obtained is again \emph{a product of single particle states}, each
rotated according to
Eqs.~(\ref{eq:single_rotated_beam},\ref{eq:single_rotated_background}).
No coherent superposition of many-particle states is formed.

It is important to emphasize that to arrive at this conclusion we
only needed to consider elementary scattering events (exchanges)
between the beam and the background. No assumptions, such as
decoherence between mass eigenstates in the background, were
necessary.

\section{One-particle evolution equations}
\label{sect:comparison}

\subsection{Evolution equation: active-active oscillations}
\label{sect:active-active}

The construction in the preceding section can be used to obtain a
differential equation describing the time evolution of a
one-particle neutrino state. We first recall that in deriving
Eq.~(\ref{eq:2beams2}) the flavor blind non-exchange interactions
were omitted. It is easy to see that to include their effect one
should add an additional term, $N_1 N_2 |e e e ...\rangle |x x x
...\rangle$, to the final state $|F_1\rangle$ in Eq.~(\ref{eq:2beams2}),
\begin{widetext}
\begin{eqnarray}
  \label{eq:2beams2_full}
 |F_1\rangle &=&
N_1 N_2(1+ |\langle e |x \rangle|^2 ) |e e e
...\rangle
|x x x ...\rangle \nonumber\\
&+& N_2 \langle\mu |x \rangle \langle x |e \rangle (|\mu e e
...\rangle+ |e \mu e ...\rangle+...)
|x x x ...\rangle\nonumber\\
&+& N_1 \langle e |x \rangle \langle y  |e \rangle |e e e
...\rangle (|y  x x ...\rangle+ |x y  x ...\rangle+...)
\nonumber\\
&+& \langle\mu |x \rangle \langle y  |e \rangle
 (|\mu e e ...\rangle+|e \mu e ...\rangle+...)
 (|y  x x ...\rangle+|x y  x ...\rangle+...).
\end{eqnarray}
\end{widetext}
Eqs.~(\ref{eq:single_rotated_beam}) and
(\ref{eq:single_rotated_background}), describing the evolution of
the one-particle neutrino states over small time $\delta t$, then
become
\begin{widetext}
\begin{eqnarray}
\label{eq:single_rotated_beam_full}
  |e'\rangle &=&
  |e \rangle + i N_2 a [1/2\times (1+|\langle x | e \rangle|^2)
|e \rangle+ \langle \mu| x \rangle \langle x  |e \rangle
|\mu\rangle],\;\;\;\\
\label{eq:single_rotated_background_full}
 |x'\rangle &=&
  |x \rangle+i N_1 a[1/2 \times  (1+|\langle e |x \rangle|^2) |x \rangle +
  \langle e |x \rangle \langle y  |e \rangle |y \rangle].\;\;\;
\end{eqnarray}
\end{widetext}

We observe that, although Eq.~(\ref{eq:single_rotated_beam_full})
was obtained for the initial states $\nu_e$ and $\nu_x$, the
derivation used only particle exchanges and did not in any way
rely on the particular choice of the initial states. Therefore, at
any time $t$, if the initial state of the beam neutrino is $\psi$
and the background neutrinos are all in the state $\phi$, the
evolution over a small time step is
\begin{eqnarray}
\label{eq:single_rotated_anyt}
  &&|\psi (t+\delta t)\rangle -|\psi (t)\rangle =
  -i \sqrt{2}G_F N_2(1-\cos\beta)/ V\delta t \times\;\;\;\nonumber\\
&&[1/2\times (1+|\langle \phi | \psi \rangle|^2) |\psi \rangle+
\langle \psi^\perp| \phi \rangle \langle \phi  |\psi \rangle
|\psi^\perp\rangle],
\end{eqnarray}
where $\psi^\perp$ is the state orthogonal to $\psi$ and we
restored the coefficients, including the angular factor.

Making use of the completeness relation
$|\psi\rangle\langle \psi|+|\psi^\perp\rangle\langle \psi^\perp|=1$,
we rewrite Eq.~(\ref{eq:single_rotated_anyt}) as
\begin{eqnarray}
\label{eq:single_rotated_anyt_2}
  &&|\psi (t+\delta t)\rangle -|\psi (t)\rangle =
  - i \sqrt{2}G_F N_2(1-\cos\beta)/V \delta t\times\;\;\;\;\;\;\nonumber\\
&&[ 1/2|\psi \rangle+| \phi \rangle \langle \phi  |\psi \rangle
 - 1/2|\langle \phi | \psi \rangle|^2
|\psi \rangle].
\end{eqnarray}
The evolution of a one-particle neutrino state is therefore
described by the following differential equation:
\begin{eqnarray}
\label{eq:oureqn}
 i \frac{d|\psi^{(i)}\rangle}{d t}&=&H_{aa}|\psi^{(i)}\rangle,\nonumber\\
H_{aa}&=&\sum_{j}\frac{\sqrt{2} G_F}{V}(1-\cos\beta^{(ij)})\times\nonumber\\
&& \left[\frac{1}{2}+|\phi^{(j)}\rangle\langle\phi^{(j)}|
-\frac{1}{2} |\langle\phi^{(j)}|\psi^{(i)}\rangle|^2 \right].
\end{eqnarray}
Here index $(i)$ refers to a given particle for which the equation
is written, and $(j)$ runs over all other neutrinos in the
ensemble. The summation over the scattering angles $\beta^{(ij)}$
was introduced to make this equation applicable to a more general
case of neutrinos propagating in different directions.


\subsection{Evolution equation: active-sterile oscillations}

The preceding analysis dealt with flavor conversions between
active neutrino states. The method developed there can be extended
to describe the conversions between active and sterile flavor
states. In doing so, one should keep in mind two important
differences between the two cases. In the active-sterile case: (i)
only active components participate in the interactions, and (ii)
the interaction amplitudes are proportional to the active content
of the neutrino states involved. Performing  calculations similar
to those in Secs. \ref{sect:exchanges} and
\ref{sect:active-active} (see the Appendix) we find that the
single-particle Hamiltonian for this case is given by
\begin{eqnarray}
\label{eq:ours_a-ster}
  H_{as}^{(i)} &=&\sum_{j}\frac{2\sqrt{2} G_F}{V}(1-\cos\beta^{(ij)})\nonumber\\
 &\times& |\langle \phi^{(j)}|e\rangle|^2 [|e\rangle\langle e|
- 1/2|\langle \psi^{(i)}|e\rangle|^2 ].
\end{eqnarray}
As before, $\psi^{(i)}$ denotes the state of the neutrino for
which the evolution equation is written and $\phi^{(j)}$
represents the flavor state of the $j$th neutrino in the
background. It is worth noting that this Hamiltonian includes both
the effects of the $t$-channel and the $u$-channel diagrams.

For comparison, the standard Hamiltonian for an active-sterile
neutrino system (the analogue of Eq.~(\ref{eq:Pantaleone})) is
\begin{eqnarray}
\label{eq:standard_a-ster}
  H^{(i)}_{as} &=&\sum_{j}\frac{2 \sqrt{2} G_F}{V}(1-\cos\beta^{(ij)})
\begin{pmatrix}
  \cos^2 \alpha^{(j)} & 0 \\
  0 & 0 \\
\end{pmatrix}\nonumber\\
&=&\sum_{j}\frac{2 \sqrt{2} G_F}{V}(1-\cos\beta^{(ij)}) |\langle
\phi^{(j)}|e\rangle|^2 |e\rangle\langle e|.\;\;\;\;
\end{eqnarray}
Here $\alpha^{(j)}$ is the mixing angle of the $j$th neutrino in
the background, $\cos^2 \alpha^{(j)}=|\langle
\phi^{(j)}|e\rangle|^2$.

\subsection{Background of antineutrinos}

We now consider flavor transformation of the neutrino beam caused
by the presence of antineutrinos in the background. For
concreteness, let us envision a modification of the thought
experiment depicted in Fig.~\ref{fig:fig1} in which the beam
contains neutrinos in a superposition of flavor states,
$\nu_z=\cos\beta\nu_e+\sin\beta\nu_\mu$, and the background
contains \emph{antineutrinos} in a flavor superposition
$\bar\nu_x=\cos\alpha\bar\nu_e+\sin\alpha\bar\nu_\mu$. To fix the
notation, let $\nu_w$ be the flavor state orthogonal to $\nu_z$,
$\langle \nu_w|\nu_z \rangle=0$ and $\bar\nu_y$ be the flavor
state orthogonal to $\bar\nu_x$, $\langle \bar\nu_y|\bar\nu_x
\rangle=0$.

Consider what elementary processes are possible in this case. As
can be easily seen, in addition to the $t$- and $u$-channel
processes, $\nu_z \bar\nu_x \rightarrow \nu_z \bar\nu_x$ and
$\nu_z \bar\nu_x \rightarrow \bar\nu_x \nu_z$, it is possible to
have $s$-channel annihilation diagrams $\nu_z \bar\nu_z
\rightarrow \nu_z \bar\nu_z, \nu_w \bar\nu_w, \bar\nu_z \nu_z,
\bar\nu_w \nu_w$. Notice that only the $\bar\nu_z$ component of
the $\bar\nu_x$ state participates in the $s$-channel processes.

The $t$-channel process is flavor-diagonal and hence does not
cause flavor conversion, just like the corresponding process in
the neutrino background. The $u$-channel process puts an
antineutrino in the beam and, therefore, cannot interfere
coherently with the incident neutrino wave. Any coherent flavor
changes, therefore, can only be due to the $s$-channel process.

Since only the $\bar\nu_z$ component of the $\bar\nu_x$ state
participates in the $s$-channel process, the beam neutrinos will
not change flavor if the background antineutrinos are in the
orthogonal flavor state, $\bar\nu_w$. While this is similar to
what was found for a background of neutrinos in the state $\nu_w$,
on the microscopical level the two cases are quite different. For
the $\nu_w$ background, the amplitude of flavor conversion for a
single elementary event is nonzero, but the conversion rate is
only proportional to $N$, because the amplitudes add up
incoherently. By comparison, for the $\bar\nu_w$ background
already the elementary amplitude vanishes and the $\nu_w$
appearance rate in the beam is strictly zero. (Instead,
$\bar\nu_w$ \emph{antineutrinos} will appear in the beam at the
rate proportional to $N$, due to the $u$-channel process.)

The elementary scattering event can be written as
\begin{eqnarray}\label{eq:elementary_antinu}
|\nu_z\bar\nu_x\rangle \longrightarrow |\nu_z\bar\nu_x\rangle -i a
\langle \bar\nu_z|\bar\nu_x \rangle
(|\nu_z\bar\nu_z\rangle+|\nu_w\bar\nu_w\rangle),
\end{eqnarray}
where the $t$-channel process as well as the processes that put an
antineutrino in the beam have been omitted. The minus sign appears
because the amplitudes for neutrino-neutrino and
neutrino-antineutrino scattering processes have opposite signs.

As before, we project the final state on the initial states and
orthogonal states,
\begin{eqnarray}
\label{eq:elementary_antinu_project}
|\nu_z\bar\nu_x\rangle &\longrightarrow&
|\nu_z\bar\nu_x\rangle -i a\langle \bar\nu_z|\bar\nu_x \rangle \times \nonumber\\
&&(\langle \bar\nu_x|\bar\nu_z \rangle |\nu_z\bar\nu_x\rangle+
\langle \bar\nu_y|\bar\nu_z \rangle |\nu_z\bar\nu_y\rangle+ \nonumber\\
&&\langle \bar\nu_x|\bar\nu_w \rangle |\nu_w\bar\nu_x\rangle+
\langle \bar\nu_y|\bar\nu_w \rangle |\nu_w\bar\nu_y\rangle).
\end{eqnarray}
The rest of the argument proceeds in complete analoguey to the case
of the neutrino background. The first three terms in parentheses
in Eq.~(\ref{eq:elementary_antinu_project}) will amplify
coherently and the last term will not. Summing over many
elementary scattering events we obtain an expression similar to
Eq.~(\ref{eq:2beams2}), which gives a one-particle evolution
equation
\begin{eqnarray}
\label{eq:oureqn_nunubar}
 i \frac{d|\psi^{(i)}\rangle}{d t}&=&H_{a\bar{a}}|\psi^{(i)}\rangle,\nonumber
 \\
 H_{a\bar{a}}&=&-\sum_{j}\frac{\sqrt{2} G_F}{V}
 (1-\cos\beta^{(ij)})\times\nonumber\\
&& \left[\frac{1}{2}+|\bar\phi^{(j)}\rangle\langle\bar\phi^{(j)}|
-\frac{1}{2} |\langle\bar\phi^{(j)}|\psi^{(i)}\rangle|^2 \right],
\end{eqnarray}
where $\bar\phi^{(j)}$ denotes the flavor state of the $j$th
antineutrino in the background and the contributions of the
$t$-channel processes have been included. Thus, the effects of the
neutrino and antineutrino backgrounds on the coherent neutrino flavor
evolution have exactly the same form (but opposite signs!), even
though at the microscopical level the two cases are different.

When both neutrinos and antineutrinos are present in the
background, their refractive effects add up linearly. Therefore,
the Hamiltonian describing the flavor evolution of a neutrino beam
equals the sum of the two contributions, Eqs. (\ref{eq:oureqn})
and (\ref{eq:oureqn_nunubar}). Further generalization to include
the effects of other matter (electrons, nucleons) and vacuum
oscillations is straightforward. Just like in the case of the usual
MSW effect, one should add the Hamiltonian terms $H_{\rm vac}$ and
$H_{\rm mat}$ to the neutrino induced Hamiltonian (see
Eq.~(\ref{eq:Pant_tot})).

\subsection{Comparison to the standard results}
\label{sect:extraterm}

We now compare the one-particle Hamiltonians we have obtained,
Eqs. (\ref{eq:oureqn}), (\ref{eq:ours_a-ster}), and
(\ref{eq:oureqn_nunubar}), to the corresponding accepted results,
Eqs. (\ref{eq:Pantaleone}) and (\ref{eq:standard_a-ster}). We see
that, while the accepted results are similar to ours, there are
important differences: in all three cases, our Hamiltonians
contain additional terms. It is important to understand both the
origin of this difference and whether it leads to any physical
consequences.

First, we would like to establish whether the presence of the
additional terms in our results affects the flavor evolution of
the neutrino system. As can be readily seen, in all three cases
the additional terms are proportional to the identity matrix in
the flavor space. The evolution equations thus have the form
\begin{eqnarray}
\label{eq:simple_oureqn}
 i\psi'=
(H_0+C(\psi,\phi)\mathbb{I})\psi,
\end{eqnarray}
where $H_0$ is the ``standard'' Hamiltonian given in
Eq.~(\ref{eq:Pantaleone}) or (\ref{eq:standard_a-ster}) and
$\mathbb{I}$ is the identity matrix. We observe that, if
$\psi_0(t)$ is the solution of the equation $i\psi'=H_0\psi$, then
$\psi_1(t)$ given by
\begin{eqnarray}
\label{eq:phase}
  \psi_1(t)=\exp\left[-i \int^t C(\psi_0(\tilde t),\phi_0(\tilde t)) d \tilde t \right]\psi_0(t)
\end{eqnarray}
is a solution of Eq.~(\ref{eq:simple_oureqn}). This means that the extra
terms give an overall shift to both energy levels, without
affecting neutrino flavor evolution.

Indeed, the physical origin of the difference in all three cases
can be traced to the part of the interaction that \emph{does not
change flavor}. For concreteness, we for a moment specialize to
the first of the three cases, the active-active conversions in the
neutrino background. It proves instructive to return to the
evolution of the beam neutrino over an infinitesimal time step. In
our case, the result is given by
Eq.~(\ref{eq:single_rotated_beam_full}), while the accepted
result, Eq.~(\ref{eq:Pantaleone}), gives
\begin{eqnarray}
\label{eq:standard_rotation}
  |e' \rangle =
  |e \rangle + i N_2 a [(1+\cos^2\alpha)
|e \rangle+ \sin\alpha\cos\alpha  |\mu\rangle].
\end{eqnarray}
The difference between the two is the factor of $1/2$, which
multiplies the state $|e\rangle$ (the flavor-preserving part) in
the brackets in Eq.~(\ref{eq:single_rotated_beam_full}). This
factor, in turn, can be traced to Eq.~(\ref{eq:2beams2_full}): it
comes from the first term in $|F_1\rangle$, which \emph{must be
split} between the beam and the background to avoid overcounting.
This is the origin of the factors of $1/2$ in
Eqs.~(\ref{eq:single_rotated_beam_full}) and
(\ref{eq:single_rotated_background_full}).

The situation is not unlike what happens in electrostatics. The
interaction energy in a system of charges is given by $1/2\sum_i
q_i \phi_i$ and the factor of $1/2$ ensures that the interaction
energy between pairs of charges is not counted twice. In our case,
the extra terms serve the same purpose, to prevent counting the
interaction energy twice. This can be seen as follows. Both
active-active and active-sterile evolution equations are
particular cases of a general case when the two flavor states have
different couplings to the $Z$ boson. As shown in the Appendix,
the evolution equation in this general case can be written in a
form (\ref{eq:oureqn_gen}):
\begin{eqnarray}
\label{eq:oureqn_gen_text}
 i \frac{d|\psi^{(i)}\rangle}{d t} = \left[H^{(i)}_0 - \frac{1}{2} \langle \psi^{(i)} | H^{(i)}_0 | \psi^{(i)}\rangle \right]
 |\psi^{(i)}\rangle~,
\end{eqnarray}
where  $H^{(i)}_0$ is the generalization of the standard
Hamiltonian \cite{germanpaper},
\begin{widetext}
\beq
H^{(i)}_0 =\sum_{j} \frac{ \sqrt{2}
G_F}{V}(1-\cos\beta^{(ij)}) \left[ G(\eta) | \phi^{(j)}
\rangle\langle \phi^{(j)} | G(\eta) + G(\eta) \langle \phi^{(j)} |
G(\eta) | \phi^{(j)} \rangle \right]~,
\eeq
\end{widetext}
with $G$ being the matrix of couplings,
\begin{eqnarray}
G(\eta)= \left(
\begin{array}{cc}
  1 & 0  \\
  0 & \eta
\end{array}
\right)~.
\end{eqnarray}

The second term in the evolution equation
(\ref{eq:oureqn_gen_text}) has the form of the expectation value
$1/2 \langle \psi^{(i)} | H^{(i)}_0 | \psi^{(i)}\rangle $. This
form makes explicit the physical meaning of this term as a
correction to avoid double counting of the energy of the system.
It is worth emphasizing that this term is \emph{always}
proportional to the identity in the flavor space, even when the
two flavor states have different couplings. Physically, this is
because the part of the interaction that needs to be split is
always the part that conserves flavor (see the Appendix).

The extra term, $1/2 \langle \psi^{(i)} | H^{(i)}_0 |
\psi^{(i)}\rangle $, depends not only on the state of the
background but also on the state of the beam itself. We therefore
caution the reader that the superposition principle, which is
always used for the MSW effect in normal matter, does not apply in
this case. For instance, suppose that a beam neutrino which is
$\nu_e$ at $t=0$ as a result of the evolution becomes a state
$\nu'$ and, similarly, a beam neutrino which is $\nu_\mu$ at $t=0$
becomes a state $\nu''$. Then, it is in general not true that the
state $\nu_x=\cos\alpha \nu_e+\sin\alpha\nu_\mu$ will become
$\cos\alpha \nu'+\sin\alpha\nu''$.

It would be incorrect to conclude that the extra terms have no
physical effect whatsoever. While they indeed do not change
neutrino flavor evolution, they do modify the absolute value of
the refraction index of a neutrino medium and, hence, at least in
principle, change the bending of a neutrino beam in a dense
neutrino medium with a density gradient. This effect is present
even if there is only one neutrino flavor in the system.

\section{More on the entangled system}
\label{sect:moreent}

We have shown that if the neutrino system initially does not
contain entangled states, such states are not formed as a result
of coherent evolution in the system. It can be argued, moreover,
that such evolution can lead to an effective loss of coherence
between entangled states. To illustrate this, let us consider a
beam in an entangled flavor state,
\begin{equation}\label{eq:ent_bg}
|{\rm ent}\rangle \equiv (|x x x ....\rangle + |y y y
....\rangle)/\sqrt{2},
\end{equation}
 (here $x $ and $ y $ are not necessarily orthogonal)
propagating in the (unentangled) background $|z z z ...\rangle$.

At time $t=0$ the expectation  value of some operator, for
example, the $\nu_\mu$ number operator, $\hat\mu=\sum
|\mu\rangle\langle\mu|$, where the sum runs over all particles in
the beam, is given by
\begin{eqnarray}
\langle {\rm ent}|\hat\mu|{\rm ent}\rangle &=& \frac{1}{2}
(\langle x x x ...|\hat\mu| x x x ...\rangle + \langle y y y
...|\hat\mu| y y y ...\rangle
\nonumber\\
&+& \langle x x x ...|\hat\mu| y y y ...\rangle + \langle y y y
...|\hat\mu| x x x ...\rangle).\;\;\;\;\;\;
\end{eqnarray}
The first two terms on the right hand side simply count the muon neutrino
content in the states $|x x x ...\rangle$ and $|y y y ...\rangle$
and the last two terms represent the effect of entanglement.

Let us consider the effects of time evolution on the expectation
value of $\hat\mu$. Each of the two terms in $|{\rm ent}\rangle$
is a product of single-particle states and according to our
earlier findings over time will remain a product of
single-particle states. Let us write the state at time $t=t_1$ as
\begin{equation}\label{eq:ent_t1}
(|x' x' x' ....\rangle|z' z' z' ...\rangle + |y'' y'' y''
....\rangle|z'' z'' z'' ...\rangle)/\sqrt{2},
\end{equation}
where $z'$, $z''$, $x'$, and $y''$ are the results of solving a
system of equations given in Eq.~(\ref{eq:oureqn}). (For example,
$z'$ and $x'$ are found by solving the equations for the initial
state $|x x x ...\rangle|z z z ...\rangle$.)

The expectation value of $\hat\mu$ in the state (\ref{eq:ent_t1})
is given by
\begin{eqnarray}\label{eq:ent_mu_t1}
&& \frac{1}{2} (\langle x' x' x' ...|\hat\mu| x' x' x' ...\rangle
+ \langle y'' y'' y'' ...|\hat\mu| y'' y'' y'' ...\rangle
\nonumber\\
&&+ \langle x' x' x' ...|\hat\mu| y'' y'' y'' ...\rangle \langle
z' z' z' ...|z'' z'' z'' ...\rangle
\nonumber\\
&&+ \langle y'' y'' y''...|\hat\mu| x' x' x' ...\rangle \langle
z'' z'' z'' ...|z' z' z'...\rangle).
\end{eqnarray}
Since the states $z'$ and $z''$ will generically be different, the
absolute value of the inner product $\langle z''|z'\rangle$ will
be $<1$. The last two terms in Eq.~(\ref{eq:ent_mu_t1}) therefore
contain suppression factors $|\langle z''|z'\rangle|^{N_2}$ and
vanish as the number of neutrinos in the background $N_2$ is taken
to infinity. As already mentioned, these terms represent the
entanglement between the two states; the system therefore behaves
as if the beam was an incoherent mixture of $| x' x' x'
...\rangle$ and $| y'' y'' y'' ...\rangle$.

Of course, rigorously speaking, the entanglement information is
not completely lost in the system. It may happen that at some time
$t$ the states $z'$ and $z''$ will be such that $|\langle
z''|z'\rangle|=1$. In this case, the entanglement effect will
reappear. We, however, regard this as an artificial arrangement
and therefore maintain that for practical purposes the coherence
is destroyed.

It is curious to note
that, as the entanglement effect
reappears, the phases of the states $|x' x' x' ....\rangle|z' z'
z' ...\rangle$ and $|y'' y'' y'' ....\rangle|z'' z'' z''
...\rangle)$ will have an effect on the expectation value of
$\hat\mu$ and hence the phases due to the additional term,
introduced in Eq.~(\ref{eq:oureqn}) to avoid overcounting of
energies, will have a physical effect.

\section{Conclusions}
\label{sect:conclusions}

In summary, we have developed a conceptually simple, physical
picture of coherent neutrino evolution in a medium of neutrinos
and have shown how coherent interference of many elementary
scattering events gives rise to the refraction phenomenon. Unlike
the case of ordinary matter, in which coherent scattering leaves
the background unchanged, in the neutrino background the
scattering events change the state of the background, thus
requiring a different type of analysis. We have found that only
part of the elementary scattering amplitude is amplified
coherently. This explains certain seemingly paradoxical results,
such as why a neutrino traveling through the medium of neutrinos
of opposite flavor does not undergo coherent flavor conversion.

We have shown that refraction  does not lead to the creation of
entangled states in a neutrino system, {\it i.e.}, if the state of
the system is initially described by a product of single-particle
states, the state remains a product of single-particle states as
the system is evolved in time. Furthermore, the evolution
effectively destroys initial entanglement in the system. It
follows that for each neutrino the result of the coherent
evolution can be described by a one-particle Schr\"odinger
equation, as was assumed {\it a priori} in the literature.

We have derived the one-particle equation
for active-active and active-sterile flavor transformation
scenarios for a neutrino in a neutrino background. We also derived
the equation for a neutrino in an antineutrino background. In all
these cases, we found that in order to avoid overcounting of the
interaction energy one has to introduce an extra term in the
evolution equations that is not present in the standard analyses.
We have proven that this extra term does not affect the flavor
evolution under normal conditions. It does, however, affect the
value of the refraction index and hence the  bending
 of a neutrino beam in a dense neutrino medium.

\acknowledgments

We thank Georg Raffelt, James Pantaleone, and John Bahcall for
stimulating discussions and for valuable comments on the
manuscript. We also acknowledge valuable comments from John
Cornwall, Wick Haxton, Maxim Perelstein, Raymond Sawyer, and
Raymond Volkas. A. F. is a Feynman Fellow at LANL. A. F. was
supported at LANL by the Department of Energy, under contract
W-7405-ENG-36, and at IAS by the Keck Foundation. C. L. was
supported by the Keck Foundation and by the National Science
Foundation grant PHY-0070928. We acknowledge the hospitality of
the Kavli Institute for Theoretical Physics, where this research
was supported in part by the National Science Foundation under
Grant No. PHY99-07949.


\section*{Appendix: the cases of active-sterile neutrinos and neutrinos with different couplings}

Let us derive the result (\ref{eq:ours_a-ster}), which refers to a
system of one active and one sterile neutrino, $\nue$ and $\nus$.
We start by considering the elementary interaction between two
neutrinos, as discussed in Sec. \ref{sec:2b}. The analogue of Eq.
(\ref{eq:pant4by4}) in this case is
\begin{widetext}
\begin{eqnarray}
  \label{eq:pant4as}
  i\frac{d}{dt}\left(
    \begin{array}{c}
      |\nu_e (\vec{k}) \nu_e(\vec{p})\rangle  \\
      |\nu_e (\vec{k}) \nu_s (\vec{p})\rangle \\
      |\nu_s (\vec{k}) \nu_e(\vec{p})\rangle \\
      |\nu_s (\vec{k}) \nu_s (\vec{p})\rangle
    \end{array}\right)
=\frac{ \sqrt{2}G_F}{V} (1-\cos\beta) \left(
\begin{array}{cccc}
  2 & 0 & 0 & 0 \\
  0 & 0 & 0 & 0 \\
  0 & 0 & 0 & 0 \\
  0 & 0 & 0 & 0
\end{array}
\right) \left(
\begin{array}{c}
      |\nu_e (\vec{k}) \nu_e(\vec{p})\rangle  \\
      |\nu_e (\vec{k}) \nu_s (\vec{p})\rangle \\
      |\nu_s (\vec{k}) \nu_e(\vec{p})\rangle \\
      |\nu_s (\vec{k}) \nu_s (\vec{p})\rangle
    \end{array}\right)~,
\end{eqnarray}
which shows that, given an initial state $|\nu_z (\vec{k}) \nu_x
(\vec{p})\rangle $,  only its active-active  component, $|\nu_e (\vec{k}) \nu_e(\vec{p})\rangle$, is affected
by the evolution:
\beq
|\nu_z (\vec{k}) \nu_x (\vec{p})\rangle
\Rightarrow  |\nu_z (\vec{k}) \nu_x (\vec{p})\rangle - i dt \frac{2
\sqrt{2}G_F}{V} (1-\cos\beta)  \langle e  | z \rangle   \langle  e
|  x \rangle  |\nu_e (\vec{k}) \nu_e(\vec{p})\rangle~.
\label{2nuas}
\eeq
\end{widetext}

Next, we apply this result to the case of two orthogonal neutrino
beams, in the spirit of what was done in Sec.
\ref{sect:exchanges}. We take neutrinos in the first and second
beams to be in the states $| z \rangle $ and $| x \rangle $,
respectively, and omit neutrino momenta for simplicity.
Similarly to Eq. (\ref{eq:2b_exch}), we get
\begin{eqnarray}
  |z z z  ...\rangle
  |x x x ...\rangle
\Rightarrow
|F\rangle =  |z z z  ...\rangle
  |x x x ...\rangle
+ i 2 a |F_1\rangle,
\end{eqnarray}
where
\begin{eqnarray}
 |F_1\rangle &=&
 \langle e  | z \rangle   \langle  e |  x \rangle   (|e z z ...z\rangle+|z e z ...z\rangle+|z z e  ...z\rangle+...)\nonumber\\
&\times&(|e x x ...x\rangle+|x e  x ...x\rangle+|x x  e
...x\rangle+...)~.\;\;\;\;
\label{2b_exch_as}
\end{eqnarray}
Notice that the effects of both the $u$-channel and $t$-channel
diagrams are included above.

One can then follow the same procedure as in Sec.
\ref{sect:exchanges} (see Eq. (\ref{eq:2beams2})) and decompose
$|F_1\rangle$ as follows:
\begin{widetext}
\begin{eqnarray}
 |F_1\rangle &=& \langle e  | z \rangle   \langle  e |  x \rangle[
N_1 N_2 \langle z |e \rangle \langle x | e \rangle | z z z
..\rangle
|x x x ...\rangle \nonumber\\
&+& N_2  \langle w |e \rangle \langle x | e \rangle (| w  z z
...\rangle+ |z w z ...\rangle+...)
|x x x ...\rangle   \nonumber\\
&+& N_1 \langle z |e \rangle  \langle y  |e \rangle  |z z z
...\rangle (|y  x x ...\rangle+ |x y  x ...\rangle+...)
\nonumber\\
&+& \langle w  |e \rangle  \langle y  |e \rangle
 (|w z z ...\rangle+|z w z ...\rangle+...)
 (|y  x x ...\rangle+|x y  x ...\rangle+...)]
 \label{eq:2beams2_as}
\end{eqnarray}
 where $| y \rangle$  and  $| w \rangle$ are the orthogonal states to $| x \rangle$
 and  $| z \rangle$ respectively ($\langle x| y \rangle=\langle z| w \rangle=0$).
The last term in Eq. (\ref{eq:2beams2_as}) is not coherently
enhanced and therefore can be dropped.  This allows us to obtain, at
first order in $a$, a factorized form:
\begin{eqnarray}
| F \rangle
\simeq   |  z^ \prime  z^ \prime  z^ \prime  ...\rangle
  | x^ \prime  x^ \prime  x^ \prime~ ...\rangle~,
\label{fact_as}
\end{eqnarray}
where
\beq
&& | z^\prime \rangle = | z \rangle +  i N_2 a |\langle e |z \rangle|^2 |\langle e | x \rangle|^2  | z \rangle   +i 2 N_2  a |\langle e | x \rangle|^2 \langle w |e \rangle \langle e |z \rangle | w \rangle \nonumber \\
&& ~~~~~ = | z \rangle +  i 2 N_2 a \langle e |z \rangle| \langle e | x \rangle|^2  | e \rangle   - i  N_2  a   |\langle e |z \rangle|^2 |\langle e | x \rangle|^2  | z \rangle \nonumber \\
&& | x^\prime \rangle = | x \rangle + i N_1 a |\langle e |z \rangle|^2 |\langle e | x \rangle|^2  | x \rangle   +i 2 N_1  a |\langle e |z \rangle|^2 \langle y |e \rangle \langle e |x \rangle | y \rangle~. \nonumber \\
&& ~~~~~ = | x \rangle +  i 2 N_1 a \langle e |x \rangle| \langle
e | z \rangle|^2  | e \rangle   - i  N_1  a   |\langle e |z
\rangle|^2 |\langle e | x \rangle|^2  | x \rangle ~.
\label{factz}
\eeq
\end{widetext}
From this a one-particle equation follows:
\begin{eqnarray}
\label{eq:oureqn_as}
 i \frac{d|\psi^{(i)}\rangle}{d t}&=&\sum_{j}\frac{2 \sqrt{2} G_F}{V}(1-\cos\beta^{(ij)})\nonumber\\
&&\times  |\langle \phi^{(j)}|e\rangle|^2 \left[ |e\rangle\langle e|
- 1/2|\langle \psi^{(i)}|e\rangle|^2 \right] |\psi^{(i)}\rangle~,\nonumber\\
\end{eqnarray}
which proves the result  (\ref{eq:ours_a-ster}).
\\

We now discuss the generalization of our findings to two active
neutrinos with different couplings to the $Z$ boson.  The study of
this case provides a unified description of the results for
active-active and active-sterile cases we have discussed so far.
Furthermore, it allows us to compare our results with the
corresponding  discussion given in ref. \cite{germanpaper}.

Let us consider two neutrino states, $\nue$ and $\nu_\rho$, and
take $\nue$ as having the ordinary Standard Model coupling, $g_e$,
with the ${\rm Z}$ boson.  The coupling of $\nu_\rho$ with the
${\rm Z}$ is defined as $g_\rho \equiv \eta g_e$. In terms of
$\eta$, the two-neutrinos equation, (\ref{eq:pant4as}), is
generalized by the replacements $\nu_s \rightarrow \nu_\rho$ and
\be \left(
\begin{array}{cccc}
2 & 0 & 0 & 0 \\
0 & 0 & 0 & 0 \\
0 & 0 & 0 & 0 \\
0 & 0 & 0 & 0
\end{array} \right)
\rightarrow B(\eta)\equiv \left(
\begin{array}{cccc}
2 & 0 & 0 & 0 \\
0 & \eta & \eta & 0 \\
0 & \eta & \eta & 0 \\
0 & 0 & 0 & 2 \eta^2 \end{array} \right)~.
\label{B}
\ee
It is manifest that the active-active (Eq.
(\ref{eq:pant4by4})) and active-sterile (Eq. (\ref{eq:pant4as})) cases
are recovered for $\eta=1$ and $\eta=0$ respectively.  The
generalization of Eq. (\ref{eq:2beams2_as}) has the form
\begin{widetext}
\begin{eqnarray}
 |F_1\rangle &=&
N_1 N_2   \langle x | \langle z | B(\eta) | z \rangle | x \rangle   | z z z
...\rangle
|x x x ...\rangle \nonumber\\
&+& N_2  \langle x | \langle w | B(\eta) | z \rangle | x \rangle  (| w  z z
...\rangle+ |z w z ...\rangle+...)
|x x x ...\rangle   \nonumber\\
&+& N_1   \langle y | \langle z | B(\eta) | z \rangle | x \rangle   |z z z
...\rangle (|y  x x ...\rangle+ |x y  x ...\rangle+...)
\nonumber\\
&+&  \langle y | \langle w | B(\eta) | z \rangle | x \rangle
 (|w z z ...\rangle+|z w z ...\rangle+...)
 (|y  x x ...\rangle+|x y  x ...\rangle+...)~.
 \label{eq:comp_gen}
\end{eqnarray}
Once the last term in Eq. (\ref{eq:comp_gen}) is neglected, as
discussed in Sec. \ref{sect:exchanges}, one gets the effective one-particle
equations:
\beq
&& | z^\prime \rangle = | z \rangle + \frac{1}{2}  i N_2 a \langle x | \langle z | B(\eta) | z \rangle | x \rangle  | z \rangle   +i  N_2  a \langle x | \langle w | B(\eta) | z \rangle | x \rangle  | w \rangle \nonumber \\
&&  | x^\prime \rangle = | x \rangle + \frac{1}{2}  i N_1 a \langle x | \langle z | B(\eta) | z \rangle | x \rangle  | x \rangle   +i  N_1  a \langle y | \langle z | B(\eta) | z \rangle | x \rangle  | y \rangle~.
\label{factz_gen}
\eeq
\end{widetext}
Notice that, as before, the factors of $1/2$ arise as a result of
splitting of the first (flavor conserving) term in
Eq.~(\ref{eq:comp_gen}).

Using the form (\ref{B}), and working out the matrix elements in
Eq. (\ref{factz_gen}), it is possible to obtain the generalization
of the result (\ref{eq:oureqn_as}). We get
\begin{eqnarray}
\label{eq:oureqn_gen}
 i \frac{d|\psi^{(i)}\rangle}{d t} = \left[H^{(i)}_0 - \frac{1}{2} \langle \psi^{(i)} | H^{(i)}_0 | \psi^{(i)}\rangle \right] |\psi^{(i)}\rangle~,
\end{eqnarray}
where  $H^{(i)}_0$ has the form
\begin{widetext}
\beq H^{(i)}_0 =\sum_{j} \frac{ \sqrt{2}
G_F}{V}(1-\cos\beta^{(ij)}) \left[ G(\eta) | \phi^{(j)}
\rangle\langle \phi^{(j)} | G(\eta) + G(\eta) \langle \phi^{(j)} |
G(\eta) | \phi^{(j)} \rangle \right]~. \label{heff} \eeq
\end{widetext}
Here the angular factors $(1-\cos\beta^{(ij)})$
have been restored to account for general orientations of the
neutrino momenta, and we use the matrix of the
couplings (normalized to $g_e$) in the $(\nue, \nu_\rho )$ basis:
\be
G(\eta)= \left(
\begin{array}{cc}
  1 & 0  \\
  0 & \eta
\end{array}
\right)~. \label{G} \ee Our result (\ref{heff}) coincides with the
one-particle Hamiltonian given in ref. \cite{germanpaper}.

As an example, we give the explicit expression of $H^{(i)}_0$ for
the case of orthogonal beams with the neutrinos in the  second
beam being all in the same state, $\phi^{(j)} = \cos \alpha \nu_e
+ \sin \alpha \nu_{\rho}$:
\begin{eqnarray}
&& H^{(i)}_0 = \frac{ \sqrt{2}G_F N_2}{V}\times\nonumber\\
&&\left(
\begin{array}{cc}
  2 \cos^2 \alpha + \eta \sin^2 \alpha  & \eta \cos \alpha \sin \alpha   \\
 \eta \cos \alpha \sin \alpha  & \eta  \cos^2 \alpha + 2 \eta^2 \sin^2 \alpha
\end{array}
\right).
\label{heff_exp}
\end{eqnarray}
It is easy to see that  Eq. (\ref{heff_exp}) reproduces  the
limiting cases of active-active ($\eta=1$) and active-sterile
($\eta=0$) neutrinos, Eqs. (\ref{eq:Pant_brief})  and
(\ref{eq:standard_a-ster}).

\bibliography{nunu}

\end{document}